\documentclass[10pt,aps,prb,twocolumn,preprintnumbers,amsmath,amssymb,floatfix,showpacs,citeautoscript,superscriptaddress]{revtex4-2} %
\usepackage[latin1]{inputenc}
\usepackage[T1]{fontenc}
\usepackage[english]{babel}
\usepackage[pdftex]{graphicx}
\usepackage{amsmath}
\usepackage{times}
\usepackage[usenames,dvipsnames,svgnames,table]{xcolor}
\usepackage[colorlinks,plainpages=false,linkcolor=blue,urlcolor=blue,citecolor=blue,pdfpagemode=UseNone]{hyperref}
\usepackage{svg}

\renewcommand{\AA}{\text{\r{A}}}

\newcommand\Vek[1]{\vec{#1}}

\begin{document}

\title
{
\boldmath
Nature of the magnetic coupling in infinite-layer nickelates versus cuprates %
}

\author{Armin Sahinovic}
\affiliation{Department of Physics and Center for Nanointegration (CENIDE), Universit\"at Duisburg-Essen, Lotharstr.~1, 47057 Duisburg, Germany}
\author{Benjamin Geisler}
\email{benjamin.geisler@ufl.edu}
\affiliation{Department of Physics and Center for Nanointegration (CENIDE), Universit\"at Duisburg-Essen, Lotharstr.~1, 47057 Duisburg, Germany}
\affiliation{Department of Physics, University of Florida, Gainesville, Florida 32611, USA}
\affiliation{Department of Materials Science and Engineering, University of Florida, Gainesville, Florida 32611, USA}
\author{Rossitza Pentcheva}
\email{rossitza.pentcheva@uni-due.de}
\affiliation{Department of Physics and Center for Nanointegration (CENIDE), Universit\"at Duisburg-Essen, Lotharstr.~1, 47057 Duisburg, Germany}

\date{\today}

\begin{abstract}
In contrast to the cuprates, where the proximity of antiferromagnetism
(AFM) and superconductivity is well established, first indications for
AFM interactions in superconducting infinite-layer nickelates were
only recently obtained.
Here, we explore, based on first-principles simulations,
the nature of the magnetic coupling in NdNiO$_2$ as a function 
of the on-site Coulomb and exchange interaction, varying the explicit 
hole doping and the treatment of the Nd $4f$ electrons. The $U$-$J$
phase diagrams for undoped nickelates and cuprates indicate $G$-type
ordering, yet show different $U$ dependency. By either Sr hole doping or
explicit treatment of the Nd $4f$ electrons, we find a transition to a Ni
$C$-type AFM ground state. We trace the effect of Sr doping back to a distinct accommodation of
the holes by the Ni versus Cu $e_g$ orbitals.
The interaction between Nd $4f$ and Ni $3d$ states stabilizes $C$-type AFM order on both sublattices.
Though spin-orbit interactions induce a band splitting near the Fermi energy,
the bad-metal state is retained even under epitaxial strain.
These results establish the distinct role of the magnetic interactions in the nickelates versus the cuprates
and suggest the former as a unique platform to investigate the relation to unconventional superconductivity.
\end{abstract}

\maketitle

\section{Introduction}

The recent discovery of superconductivity in Sr-doped NdNiO$_2$, PrNiO$_2$, and LaNiO$_2$ films grown on 
SrTiO$_3$(001) (STO)~\cite{Li-Supercond-Inf-NNO-STO:19, Li-Supercond-Dome-Inf-NNO-STO:20, Osada-PrNiO2-SC:20, SC-WO-REmag-Osada:21, Zeng_LaCaNiO2:22}
initiated considerable interest in infinite-layer ($AB$O$_2$) nickelates~\cite{Nomura-Inf-NNO:19, JiangZhong-InfNickelates:19, NiO2-Mottness:20, Botana-Inf-Nickelates:19, Lechermann-Inf:20, trace_of_AFM_Liu:20, Chen_magnetism_in_doped_2022, NiO2-interface-Geisler:20, NiO2-interface-Geisler:21, NiO2-interface-Geisler-Hwang:22, Model_Construction:20, sahinovic_active:21, sahinovic_quantifying:22, Topotactic_Hydrogen_Si:20, DMFT-NNO-chen:22, Geisler-VO-LNOLAO:22, Geisler-Rashba-NNOSTOKTO:23}.
Specifically, their formal Ni$^{1+}$ ($3d^9$) valence state renders them close to the cuprates~\cite{NiO2-CuO2-analog:99},
with a single hole in the $d_{x^2-y^2}$ orbital.
Simultaneously, a number of differences between the two materials classes have been pointed out~\cite{NdNiO2-ni-is-not-cu:04, Botana-Inf-Nickelates:19}.
A controversial aspect is whether the electronic properties are dominated
by single-band physics, as typical for the cuprates~\cite{Zhang-Rice-Singlett:88},
or if they are rather of multi-orbital nature,
e.g., affected by the Nd-$5d$- and Ni-$3d$-derived self-doping pockets~\cite{Botana-Inf-Nickelates:19, Lechermann-Inf:20, NdNiO2-Hepting-study:20, NdNiO2-Multiorbital:20, NiO2-one-band:20, NNO-SelfDopingDesign-d9-Arita:20, NiO2-holes-in-Ni:21, NdNiO2-lanthanide-trends:21}.
Another important distinction consists in the presence of rare-earth $4f$ electrons in the nickelates, with the notable exception of LaNiO$_2$~\cite{Choi-Lee-Pickett-4fNNO:20, Bandyopadhyay-4f-in-NdNiO2:22}.
Most prominently, superconductivity remained elusive in the respective bulk compounds so far~\cite{Li-NoSCinBulkDopedNNO:19, Wang-NoSCinBulkDopedNNO:20},
which raised a question about the polar interface to the substrate~\cite{NiO2-interface-Geisler:20, NiO2-interface-Geisler:21},
with recent indications of a complex interface composition~\cite{NiO2-interface-Geisler-Hwang:22}.
Finally, the pairing mechanism is not yet fully understood~\cite{ARXIV_nodal_sym:22, ARXIV_pairing_sym:22},
and even the notion of nickelate superconductivity being unconventional has recently been challenged~\cite{ARXIV_two_gap_SC_Louie:22}.
Therefore, understanding the nature of the magnetic interactions in these compounds is of fundamental relevance.

Experimental evidence of antiferromagnetic (AFM) long-range order characteristic of the cuprates is absent in the infinite-layer nickelates~\cite{Li-Supercond-Inf-NNO-STO:19}.
Intriguingly, an intrinsic short-range AFM ground state~\cite{NiO2-intrinsic-magnetism:22}
and magnetic excitations consistent with cuprate-like AFM interactions~\cite{NdNiO2-magnetic-excitations:21}
have recently been identified in NdNiO$_2$ films on SrTiO$_3$(001). %
Another topical work reports differences in the magnitude and anisotropy of the superconducting upper critical field
in NdNiO$_2$ versus PrNiO$_2$ and LaNiO$_2$ films on SrTiO$_3$(001),
and traces the distinct polar and azimuthal angle-dependent magnetoresistance of NdNiO$_2$
to the magnetic contribution of the finite Nd$^{3+}$ $4f$ moment~\cite{ARXIV_magneto_resistance:22}.
Furthermore, measurements of the London penetration depth 
suggest qualitative differences in the superfluid density
between these three compounds~\cite{ARXIV_nodal_sym:22}.

In the light of these new observations, we provide a comprehensive picture of the magnetic interactions in infinite-layer nickelates versus cuprates by performing first-principles simulations
including a Coulomb repulsion term.
We systematically and consecutively vary a number of control parameters,
i.e.,
the on-site Coulomb and exchange interaction, 
the explicit hole doping, and the treatment of the Nd $4f$ electrons.
We compile $U$-$J$ phase diagrams for nickelates and cuprates,
which, in the undoped case, indicate $G$-type AFM ordering, yet with an opposite $U$ dependence of the magnetic coupling.
Sr doping leads to a transition to a Ni $C$-type AFM ground state, which is attributed to a distinct response
of the Ni versus Cu $e_g$ orbitals to the hole doping. 
Likewise, also the explicit treatment of the Nd $4f$ electrons in NdNiO$_2$ leads to a $C$-type AFM order on both sublattices due to the coupling between Nd $4f$ and Ni $3d$ states,
marking a fundamental difference to LaNiO$_2$.
Even though spin-orbit interactions induce a band splitting near the Fermi energy,
the bad-metal state is retained even under epitaxial strain in the range of $3.79~\AA$ to $3.94~\AA$ as imposed by different substrates.
These results promote infinite-layer nickelates as a unique platform to gain a more profound understanding of unconventional superconductivity
and its relation to the underlying magnetic interactions.
\section{Methodology}

We performed first-principles simulations in the framework
\filbreak
of spin-polarized density functional theory (DFT)~\cite{KoSh65}
as implemented in Quantum ESPRESSO (QE)~\cite{PWSCF} 
and the \textit{Vienna Ab initio Simulation Package} (VASP)~\cite{kresse1996a, kresse1996b},
using the generalized gradient approximation as parametrized by Perdew, Burke, and Ernzerhof~\cite{PeBu96}
and, for reference calculations, the 'strongly constrained and appropriately normed' (SCAN) semilocal exchange-correlation functional~\cite{SCAN:2015}.
Static correlation effects were considered within the DFT$+U$ formalism~\cite{LiechtensteinAnisimov:95} %
employing, if not specified otherwise, $U_\text{Ni, Cu}^d = 5$~eV and $J_\text{Ni, Cu}^d = 1$~eV,
similar to previous work~\cite{ZhongKosterKelly:12, Liu-NNO:13, Geisler-LNOSTO:17, WrobelGeisler:18, GeislerPentcheva-LNOLAO:18, GeislerPentcheva-LNOLAO-Resonances:19, Botana-Inf-Nickelates:19}.
The wave functions and density were expanded into plane waves up to cutoff energies of $45$ and $350$~Ry in QE, respectively,
and $500$~eV in VASP. %
Ultrasoft pseudopotentials~\cite{Vanderbilt:1990},
as successfully employed in previous work~\cite{GeislerPopescu:14, Geisler-Heusler:15, GeislerFePcHSi:19},
were used in conjunction with projector augmented wave datasets~\cite{PAW:94}.
The impact of the Nd $4f$ states is analyzed by comparing results obtained by treating them explicitly (applying $U_\text{Nd}^f = 8$~eV and $J_\text{Nd}^f = 1$~eV~\cite{Choi-Lee-Pickett-4fNNO:20})
to those rendered by the frozen-core approximation.
Spin-orbit calculations were performed in a fully self-consistent approach by using the VASP code.

Infinite-layer NdNiO$_2$, LaNiO$_2$, and CaCuO$_2$ are modeled in $\sqrt{2}a \times \sqrt{2}a \times 2c$ supercells that feature four inequivalent transition-metal sites,
adopting the following lattice parameters: %
$a = 3.92$, $c = 3.28~\AA$ (NdNiO$_2$~\cite{Hayward:03, Botana-Inf-Nickelates:19, Nomura-Inf-NNO:19}),
$a = 3.96$, $c = 3.37~\AA$ (LaNiO$_2$~\cite{Hayward:99, Botana-Inf-Nickelates:19}), and
$a = 3.86$, $c = 3.20~\AA$ (CaCuO$_2$~\cite{Botana-Inf-Nickelates:19}).
Epitaxial strain was considered by setting the lateral lattice parameter to
$a_\text{LAO} = 3.79$ (LaAlO$_3$),
$a_\text{STO} = 3.905$ (SrTiO$_3$), and
$a_\text{DSO} = 3.94~\AA$ (DyScO$_3$), %
and subsequently optimizing the cell height $c$. %
We used a $12 \times 12 \times 10$ Monkhorst-Pack $\Vek{k}$-point grid~\cite{MoPa76}
and $5$~mRy Methfessel-Paxton smearing~\cite{MePa89} to sample the Brillouin zone.

\section{\boldmath Magnetic coupling as a function of the on-site Coulomb and exchange interactions} %

\begin{figure}[t]
    \includegraphics[width=\linewidth]{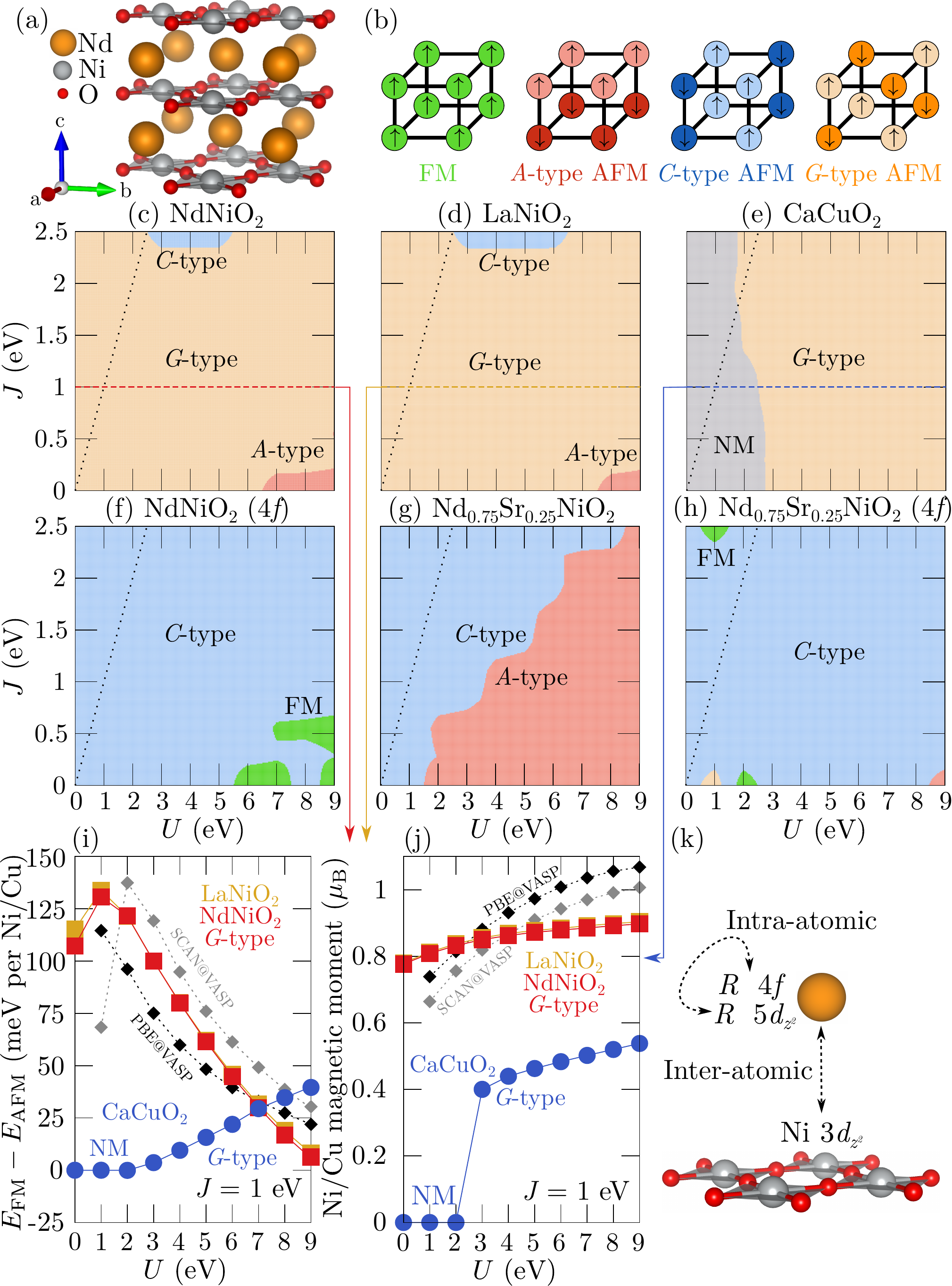}
    \caption{
        Modeling the $AB$O$_2$ infinite-layer structure in $\sqrt{2}a \times \sqrt{2}a \times 2c$ supercells~(a) permits us to consider
        four different magnetic orderings~(b) independently at the $A$- and $B$-site sublattices.
        (c-e)~Magnetic phase diagrams for NdNiO$_2$, LaNiO$_2$, and CaCuO$_2$ 
        show the ground-state order at the Ni/Cu sublattice for a range of
        $U$ and $J$ values (the dotted line marking $U=J$),
        which suggests a profound $G$-type order in all three compounds.
        Closer inspection unravels a fundamentally distinct $U$ dependence of the Ni/Cu magnetic coupling in NdNiO$_2$ and LaNiO$_2$ versus CaCuO$_2$, displayed in panel~(i).
        The evolution of the corresponding Ni/Cu magnetic moments is shown in panel~(j).
        The black and grey dashed lines in panels~(i) and~(j) 
        confirm these trends by using the VASP code and the SCAN exchange-correlation functional.
        (f)~Surprisingly, the explicit treatment of the Nd $4f$ electrons (so far FM ordered) reveals an actual $C$-type magnetic ground state of the Ni ions in NdNiO$_2$, demonstrating a strong coupling of the sublattices.
        A possible mechanism is illustrated in panel~(k).
        (g,h)~The explicitly hole-doped Nd$_{0.75}$Sr$_{0.25}$NiO$_2$ exhibits $C$-type AFM order as well, which is additionally stabilized by a full consideration of the Nd $4f$ states.
        }
    \label{fig:phasediagrams}
\end{figure}

The combination of four different spin alignments on both the Nd and Ni sublattices [FM, $A$-type, $C$-type, $G$-type; Fig.~\ref{fig:phasediagrams}(a,b)]
gives rise to 16 distinct magnetic phases in total.
By comparing their relative stability, we establish the magnetic ground state of NdNiO$_2$ to be Ni $C$-type, Nd $C$-type AFM,
consistently rendered by VASP and QE,
which confirms earlier findings~\cite{Choi-Lee-Pickett-4fNNO:20}.
In order to provide a more profound understanding of the nature of the magnetic coupling,
we begin by consecutively disentangling the individual contributions of $U$ and $J$ in infinite-layer nickelates versus cuprates,
as well as the influence of the $4f$ states and the impact of explicit Sr hole doping.

Figure~\ref{fig:phasediagrams}(c-e) compares the magnetic phase diagrams for NdNiO$_2$, LaNiO$_2$, and CaCuO$_2$,
which display the predicted ground-state order at the Ni/Cu sites as a function of the corresponding $U$ and $J$ values.
The phase diagrams have been obtained by performing a large number of individual DFT$+U$ simulations on a dense ($U,J$) grid
for all magnetic orders shown in Fig.~\ref{fig:phasediagrams}(b).
NdNiO$_2$ (with the common frozen-core treatment of the $4f$ electrons) and LaNiO$_2$ ($4f^0$)
exhibit $G$-type AFM order over a large parameter range,
similar to the established antiferromagnet CaCuO$_2$.
The emergence of an AFM ground state in NdNiO$_2$ is in agreement with the very recent observation of AFM interactions in infinite-layer nickelate films on SrTiO$_3$(001)~\cite{NdNiO2-magnetic-excitations:21, NiO2-intrinsic-magnetism:22},
as well as theoretical DFT$+U$ investigations treating the $4f$ electrons as core states, resulting in a $G$-type AFM ground state~\cite{trace_of_AFM_Liu:20}.

Simultaneously, a more detailed inspection unravels that the energy difference between the FM/NM state and the energetically lowest AFM phase displays an inherently different $U$ dependence [Fig.~\ref{fig:phasediagrams}(i)].
Surprisingly, NdNiO$_2$ and LaNiO$_2$ display pronounced AFM order even for a vanishing Coulomb repulsion term,
which is found to destabilize with increasing~$U$.
In contrast, the AFM phase in CaCuO$_2$ emerges at finite $U > 2$~eV and then continuously stabilizes with increasing~$U$.
A crossover of the curves is observed at $U = 7$~eV.
The local Ni and Cu magnetic moments consistently increase with~$U$ [Fig.~\ref{fig:phasediagrams}(j)].
While quantitative details may depend on the technical implementation of the DFT$+U$ formalism, 
the general trends of the magnetic coupling energies and the local Ni magnetic moments agree nicely between PBE$+U$ and SCAN$+U$ and between QE and VASP. 
Furthermore, we confirmed the predicted $G$-type ground state for NdNiO$_2$ (without explicit consideration of $4f$ states) and for LaNiO$_2$
by using VASP in conjunction with the PBE$+U$ and SCAN$+U$ exchange-correlation functionals.
A recent work employing the SCAN exchange-correlation functional indicates a $C$-type ground state for LaNiO$_2$, %
albeit with a relatively small energy difference to the $G$-type phase~\cite{LNO_SCAN_Zhang:21}
that may be susceptible to further details of the simulations.
In any case, %
the clear qualitative differences in magnetic coupling energies and local magnetic moments exemplifies the very distinct physics in the two infinite-layer families and can be traced back to the 
distinct Ni $3d$-O $2p$ versus Cu $3d$-O $2p$ hybridization %
\cite{Botana-Inf-Nickelates:19, NiO2-interface-Geisler:21},
which also leads to a higher charge-transfer energy in the nickelates and more Mott-like physics~\cite{NiO2-Mottness:20} than in the cuprates.

\begin{figure}[t]
    \includegraphics[width=\linewidth]{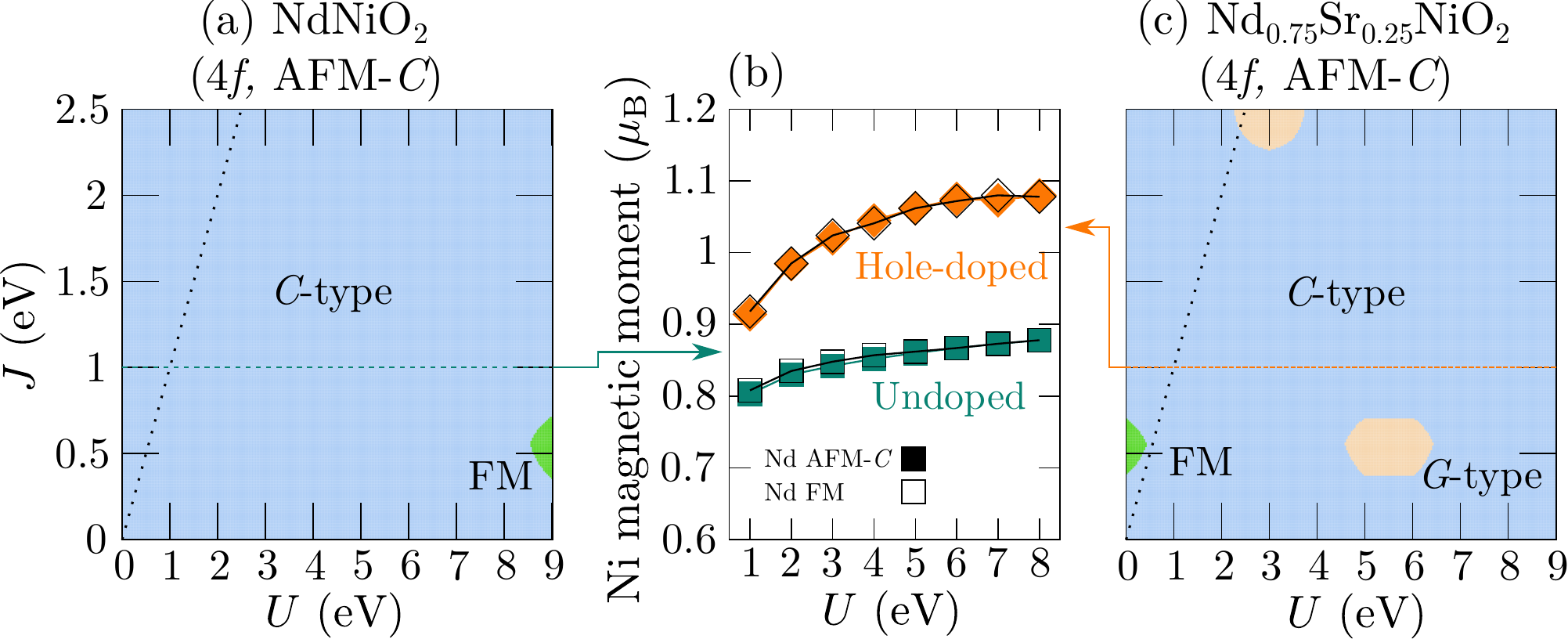}
    \caption{(a)~Magnetic phase diagram of NdNiO$_2$ with explicit treatment of the Nd~$4f$ electrons, now in their ground-state AFM $C$-type order. Comparison with Fig.~\ref{fig:phasediagrams}(f) reveals that the AFM $C$-type ground state at the Ni sites is preserved.
    The same is observed for the explicitly hole-doped Ni system achieved by doubling the supercell in vertical direction [panel~(c); see also Fig.~\ref{fig:phasediagrams}(h)].
    (b) The site-averaged Ni magnetic moments of undoped and hole-doped NdNiO$_2$ monotonically increase with~$U$;
    hole doping results in a sizeable additional increase of their magnitude.
    In contrast, they are quasi independent of the Nd magnetic order.
    }
    \label{fig:Nd-AFM-C}
\end{figure}

Next, we investigate the interplay between the Ni and Nd sublattices.
To this extent, the $4f$ electrons are now explicitly treated and, for the moment, considered to be FM ordered. %
Interestingly, we observe a transition of the Ni-site magnetic order 
to $C$-type AFM when the Nd $4f$ electrons are explicitly included [Fig.~\ref{fig:phasediagrams}(f)], regardless of an adopted FM or ground-state $C$-type AFM order of the Nd ions [Fig.~\ref{fig:Nd-AFM-C}(a)]. 
The identified Ni $C$-type AFM ground state is consistent with earlier work that considered the Nd~$4f$ states~\cite{LNO_SCAN_Zhang:21, Choi-Lee-Pickett-4fNNO:20, zhang_structural_2022}.
This Ni $3d$-Nd $4f$ coupling
can be interpreted by an indirect mechanism
via an \textit{intra-atomic} interaction between the Nd $4f$ states
localized $\sim 6.5$~eV below the Fermi energy
(see the discussion of the band structures below; also Ref.~\cite{Choi-Lee-Pickett-4fNNO:20})
and the Nd $5d$ states at the Fermi level,
in conjunction with the \textit{inter-atomic} Nd $5d$-Ni $3d$ hybridization mediated by itinerant electrons,
as illustrated in Fig.~\ref{fig:phasediagrams}(k).
Thereby, the presence of finite $4f$ moments leads to a considerably different Ni magnetic coupling in LaNiO$_2$ versus NdNiO$_2$.

The role of the rare-earth $4f$ electrons in the pairing mechanism remains so far ambivalent.
Both LaNiO$_2$ ($4f^0$) and NdNiO$_2$ ($4f^3$) exhibit a superconducting phase in film geometry,
with a superconducting dome of slightly different shape as a function of hole doping~\cite{SC-WO-REmag-Osada:21}.
We therefore conclude that the discussed Nd-$4f$-Nd-$5d$-Ni-$3d$ coupling mechanism is not critical for the emergence of superconductivity.
Moreover, contrary to NdNiO$_2$ and PrNiO$_2$ ($4f^2$),
LaNiO$_2$ shows a tendency towards superconductivity even in the undoped case~\cite{SC-WO-REmag-Osada:21},
which may point to a rather impeding nature of the rare-earth $4f$ states. 
Simultaneously, differences in the magnitude and anisotropy of the superconducting upper critical field~$H_{c2}$ have been reported very recently
in NdNiO$_2$ versus PrNiO$_2$ and LaNiO$_2$ films on SrTiO$_3$(001),
and NdNiO$_2$ features a unique polar and azimuthal angle-dependent magnetoresistance
that can be understood from the magnetic contribution of the finite Nd$^{3+}$ $4f$ moment,
which is absent for La$^{3+}$ as well as for Pr$^{3+}$ if a singlet state is adopted~\cite{ARXIV_magneto_resistance:22}.

Finally, we explore how the phase diagrams change under the presence of 25\% Sr doping,
a representative value that marks the transition of the superconducting phase to the overdoped regime~\cite{Li-Supercond-Dome-Inf-NNO-STO:20}.
The additional holes unambiguously modify the magnetic coupling, which leads to the emergence of a competing $A$-type AFM phase at higher $U-J$ values [Fig.~\ref{fig:phasediagrams}(g)],
characterized by a parallel spin alignment in the basal plane.
Intriguingly, the explicit treatment of the $4f$ states quenches this phase,
and the $C$-type AFM order at the Ni sites stabilizes across the entire phase diagram of Nd$_{0.75}$Sr$_{0.25}$NiO$_2$ [Fig.~\ref{fig:phasediagrams}(h)].
A largely identical picture is obtained for $C$-type AFM order at the Nd sublattice instead of FM order [Fig.~\ref{fig:Nd-AFM-C}(c)].  
Notably, we doubled the supercell in this case to obtain a fully compensated AFM order at the Nd sites,
which otherwise would be disturbed by the substituted Sr ions.

\begin{figure}[t]
    \includegraphics[width=\linewidth]{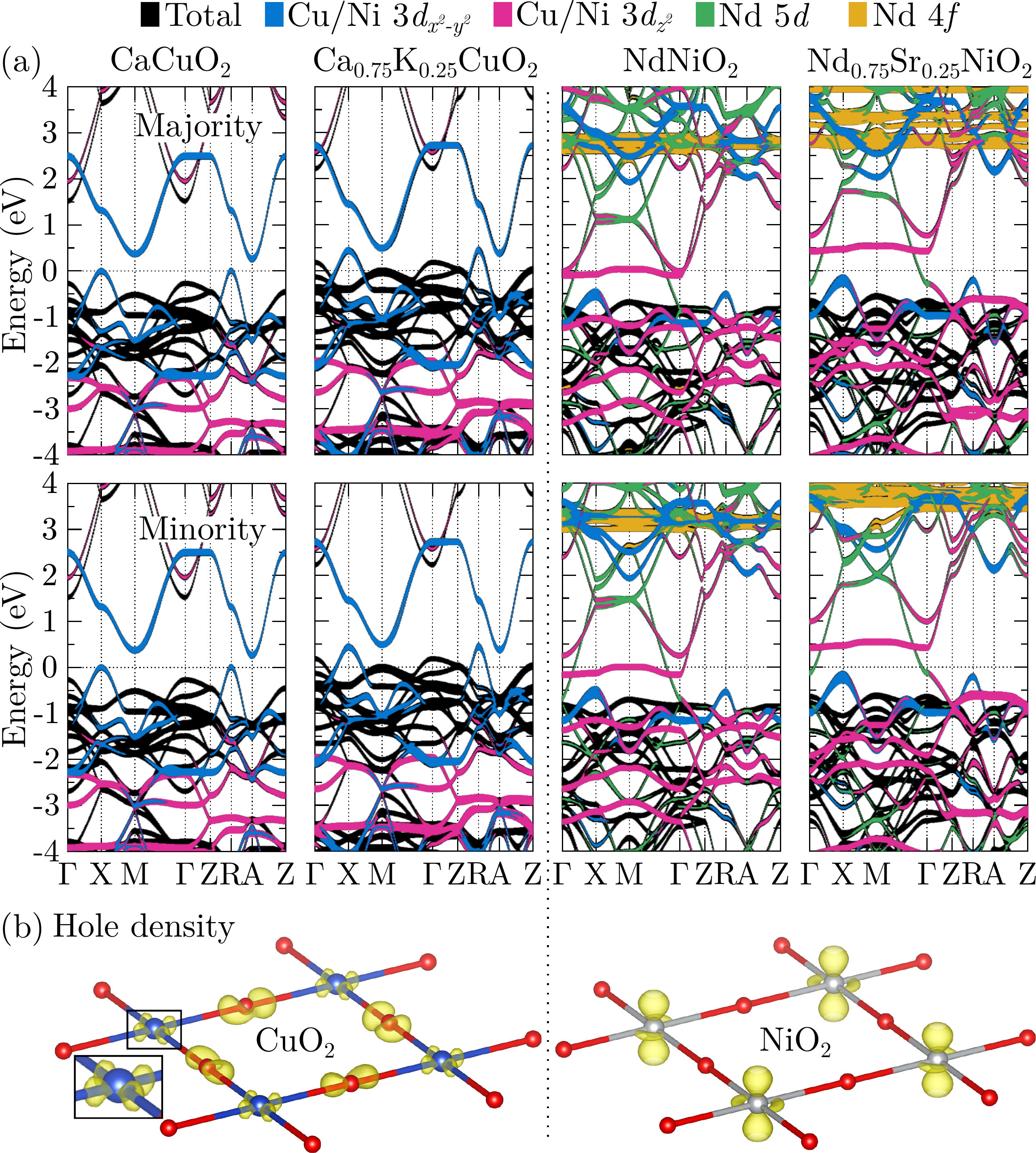}
    \caption{(a)~Spin-resolved band structure of undoped and hole-doped CaCuO$_2$ ($G$-type; left) and NdNiO$_2$ (Ni $C$-type, Nd FM; right).
        The Cu/Ni 3$d_{x^2-y^2}$, Cu/Ni 3$d_{z^2}$, Nd 5$d$, and Nd 4$f$ orbitals are depicted by blue, pink, green, and orange, respectively.
        In stark contrast to the infinite-layer cuprates, where the hole doping strongly affects the oxygen $2p_x, 2p_y$ system (represented by the black valence bands) and slightly reduces the Cu magnetic moment via the $3d_{x^2-y^2}$ orbital,
        exclusively the Ni states accommodate the doped holes in the infinite-layer nickelates,
        resulting in a depleted Ni-$3d_{z^2}$-derived flat band
        and significantly enhanced Ni magnetic moments.
        (b)~This is also reflected by the density difference plots,
        which visualize the areas of increased hole density due to doping
        (isosurface value $n_c = 0.006$~a.u.$^{-3}$ for CaCuO$_2$ and $0.012$~a.u.$^{-3}$ for NdNiO$_2$).
    }
    \label{fig:Bands}
\end{figure}

\section{Distinct accommodation of explicit hole doping in nickelates versus cuprates} %

In order to rationalize the qualitative differences in the phase diagrams discussed above,
we now analyze and compare the effect of hole doping on the electronic structure in nickelates versus cuprates.
It has been proposed that the $S=1/2$ spin state of the undoped nickelate (Ni$^{1+}$)
turns into $S=1$ upon hole doping (Ni$^{2+}$),
which is in conflict with a cuprate-like superconductivity mechanism~\cite{NiO2-Mottness:20}.
This is related to the distinct charge-transfer energy of nickelates versus cuprates~\cite{NiO2-holes-in-Ni:21, NdNiO2-Multiorbital:20},
which suggests a rather Mott-like physics and a different accommodation of the doped holes.
Another question is how the self-doping pockets, which are a characteristic property of the nickelates~\cite{NdNiO2-ni-is-not-cu:04, Botana-Inf-Nickelates:19},
respond to hole doping of the bulk compound.

In order to shed some light into these fundamental aspects, we model the hole doping in nickelates and cuprates
by explicitly substituting 25\% Nd by Sr in NdNiO$_2$ and, to mimick a comparable scenario, 25\% Ca by K in CaCuO$_2$.
The distinct magnetic ground state leads to a highly different electronic structure in the two compounds
despite the formally equivalent $3d^9$ configuration [Fig.~\ref{fig:Bands}(a)].
For the undoped infinite-layer cuprate,
the singly occupied Cu $3d_{x^2-y^2}$ states that constitute the Fermi surface in the nonmagnetic case~\cite{Botana-Inf-Nickelates:19}
also delimit the emerging band gap in the $G$-type AFM phase,
with Cu magnetic moments of $\pm 0.46~\mu_\text{B}$ [see also Fig.~\ref{fig:phasediagrams}(j)].
In contrast, the Cu $3d_{z^2}$ states are fully occupied due to the tetragonal crystal-field splitting. %
Consequently, the doped holes in Ca$_{0.75}$K$_{0.25}$CuO$_2$
are exclusively accommodated by the Cu $3d_{x^2-y^2}$ and O $2p_\sigma$ states in the valence band [Fig.~\ref{fig:Bands}(b)],
concomitantly lowering the Cu magnetic moments to $\pm 0.37~\mu_\text{B}$.
The noninteger Cu magnetic moment (also observed in La$_2$CuO$_4$~\cite{WrobelGeisler:18})
and its change upon doping that is apparently inconsistent with $0.25$ holes per Cu ion
underline the strong hybridization with the O $2p$ states.

In sharp contrast, the undoped infinite-layer nickelate 
exhibits $C$-type AFM order [Fig.~\ref{fig:phasediagrams}(f)] with Ni magnetic moments of $\pm 0.86~\mu_\text{B}$ [Figs.~\ref{fig:phasediagrams}(j), \ref{fig:Nd-AFM-C}(b)],
and is metallic [Fig.~\ref{fig:Bands}(a)],
in agreement with recent experimental observations~\cite{Li-Supercond-Inf-NNO-STO:19, Li-Supercond-Dome-Inf-NNO-STO:20}.
The energy difference between the occupied and the empty $3d_{x^2-y^2}$ states is much larger in NdNiO$_2$ ($\sim 2.5$~eV) than in CaCuO$_2$ ($\sim 0.4$~eV).
The active states at the Fermi level are of Ni $3d_{z^2}$ character,
either in the form of a depleted electron pocket at the $\Gamma$ point directly above the Fermi energy
that self-dopes the bulk compound in the nonmagnetic case~\cite{Botana-Inf-Nickelates:19, Lechermann-Inf:20},
or a flat band that crosses the Fermi level along $X$-$M$ and $M$-$\Gamma$.
Additionally, highly dispersive Nd $5d$ states cross the Fermi energy,
which constitute the 'interstitial' electron pocket at the $A$ point in the nonmagnetic case~\cite{Nomura-Inf-NNO:19, Model_Construction:20}.
Their mutual presence at the Fermi surface couples the Nd $5d$ and Ni $3d$ states magnetically (Fig.~\ref{fig:phasediagrams})
and diminishes the cuprate-like two-dimensional electronic character of the $B$O$_2$ planes.

Hole doping lowers the Fermi level in Nd$_{0.75}$Sr$_{0.25}$NiO$_2$ closer to the Ni $3d_{x^2-y^2}$ and O $2p$ states,
but even at 25\% Sr substitution they remain fully occupied,
which constitutes a major difference to the cuprates.
Instead, the Nd $5d$ and Ni $3d_{z^2}$ states are partially depleted [Fig.~\ref{fig:Bands}(b)], %
which significantly increases the Ni magnetic moments to $\pm~1.06~\mu_\text{B}$ [see also Fig.~\ref{fig:Nd-AFM-C}(b)],
consistent with $\sim 0.25$ holes per formula unit
and fundamentally different from the cuprates.
This demonstrates that the picture of charge-transfer (cuprates) versus Mott physics (nickelates) is appropriate,
even though the change in spin state under realistic doping conditions is more moderate than suggested earlier~\cite{NiO2-Mottness:20}.

The Ni-$3d_{z^2}$-derived flat band at the Fermi energy,
which is clearly absent in the cuprates,
suggests a bond instability at lower temperatures~\cite{NdNiO2-flatband:20} under appropriately chosen $U$ and $J$ parameters and exchange-correlation functional.
In the present study, neither a checkerboard nor a stripe-ordered bond disproportion could be stabilized in the undoped system,
not even in large $2\sqrt{2} \times 2\sqrt{2} \times 2$ unit cells including spin-orbit effects and Nd $4f$ electrons.
Moreover, finite Sr doping pushes the flat band to $\sim 0.5$~eV [Fig.~\ref{fig:Bands}(a)], which renders this scenario even more unlikely.

These results underpin the notion of a pairing mechanism in bulk nickelates
that deviates to a certain extent from that of the cuprates.
If superconductivity is unconventional and mediated by spin fluctuations,
it is reasonable to expect a modulation due to the distinct electronic and magnetic properties
and response to doping.
Notably, the situation may be different in film geometry due to a pronounced electronic reconstruction~\cite{NiO2-interface-Geisler:20, NiO2-interface-Geisler:21, NiO2-interface-Geisler-Hwang:22}.

\begin{figure}
    \includegraphics[width=\linewidth]{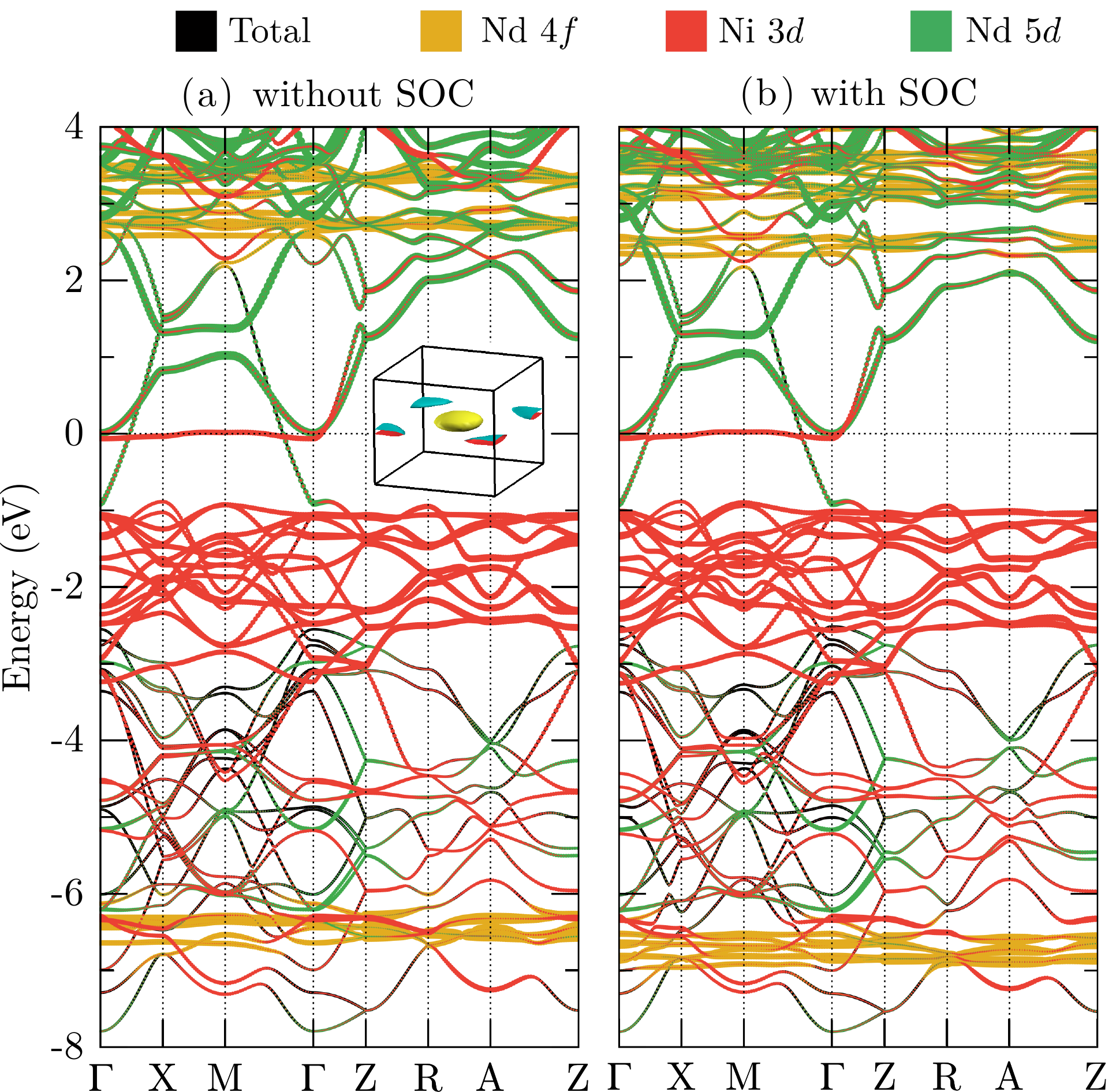}
    \caption{Band structure of NdNiO$_2$ in the Ni $C$-type, Nd $C$-type AFM ground state from (a)~DFT$+U$ and (b)~DFT$+U+$SOC.
        Orange, red, and green bands indicate Nd $4f$, Ni $3d$, and Nd $5d$ orbital character, respectively.
        The corresponding Fermi surface is localized in the $k_x, k_y$ plane (inset).
    }
    \label{fig:SOC}
\end{figure}

\begin{figure*}
    \includegraphics[width=\textwidth]{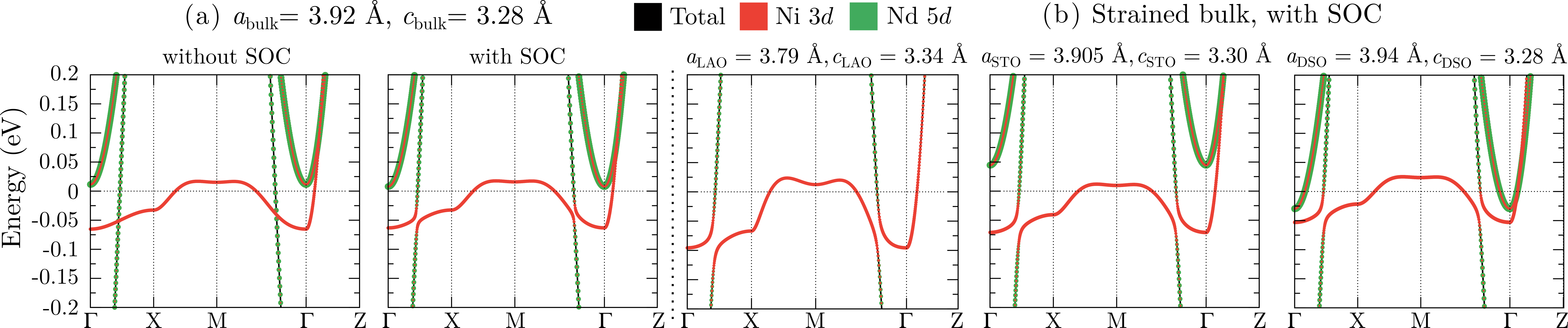}
    \caption{Strain-dependent band structure of fully $C$-type AFM NdNiO$_2$ around the Fermi level. 
        (a)~Spin-orbit coupling results in hybridization and avoided crossings between Ni-$3d$-derived (red) and Nd-$5d$-derived (green) bands.
        (b)~Nevertheless, neither compressive ($a_\text{LAO}$, $a_\text{STO}$) nor tensile epitaxial strain ($a_\text{DSO}$) drive a metal-to-insulator transition.
        Instead, a weakly conducting metal is retained.
        Merely a strong response of the normally depleted Nd-$5d$-derived electron pocket at the $\Gamma$ point can be observed.
        }
        \label{fig:SOCStrain}
\end{figure*}

\section{Impact of spin-orbit coupling and strain}

Figure~\ref{fig:SOC}(a) shows the DFT$+U$ band structure of bulk NdNiO$_2$ in the fully $C$-type AFM ground state,
obtained by using the VASP code.
The three occupied Nd $4f$ bands can be observed at $\sim -6.5$~eV
and give rise to a Nd magnetic moment of $\pm 3~\mu_\text{B}$.
We identified the corresponding orbitals to be $4f_{y(3x^2-y^2)}$, $4f_{z(x^2-y^2)}$, and $4f_{xz^2}$
($m_l = -3, -2, +1$),
which perfectly matches earlier results obtained with the linearized augmented plane-wave method~\cite{Choi-Lee-Pickett-4fNNO:20}.
The Fermi surface is sharply localized in the $k_x, k_y$ plane
and features a Nd $5d$ electron pocket at the $\Gamma$ point
in conjunction with a hole pocket at the $M$ point that originates from the Ni $3d$ flat band.
These properties are a unique feature of the nickelate and absent in CaCuO$_2$.

The inclusion of spin-orbit coupling (SOC) results in an overall similar band structure [Fig.~\ref{fig:SOC}(b)].
The most pronounced differences can be observed for the Nd $5d$ states (e.g., around $1$~eV)
and in modifications of the Nd $4f$ bands (around $-6.5$ and above $2.5$~eV).
The latter is paralleled by a reduction of the Nd magnetic moment to $\sim 1.8~\mu_\text{B}$
and the concomitant emergence of a large Nd orbital moment of $\sim 3.8~\mu_\text{B}$.
The reduced Nd magnetic moment is closer to the notion of a Kramer's doublet state adopted by Nd$^{3+}$~\cite{ARXIV_magneto_resistance:22}
than to a full polarization of the spins according to Hund's rule.
In contrast, the Ni magnetic moment remains unchanged [$0.98~\mu_\text{B}$ in VASP; see also Fig.~\ref{fig:phasediagrams}(j)].

Closer inspection of the band structure around the Fermi level [Fig.~\ref{fig:SOCStrain}(a)]
unveils that crossings between the Ni $3d$ flat band and the Nd $5d$ bands are lifted
due to a SOC-induced hybridization between the orbitals, which is otherwise suppressed by symmetry.
Motivated by this observation,
we probe in Fig.~\ref{fig:SOCStrain}(b) the response of the rehybridized pockets to compressive ($a_\text{LAO}$, $a_\text{STO}$) as well as tensile epitaxial strain ($a_\text{DSO}$).
Intriguingly, in neither case a metal-to-insulator transition is achieved, and the experimentally observed weakly conducting metallic phase~\cite{Li-Supercond-Inf-NNO-STO:19, Li-Supercond-Dome-Inf-NNO-STO:20} is retained.
Nevertheless, we observe a charge transfer between the different pockets due to strain.
Compressive strain promotes the occupation of the Ni~$3d_{z^2}$-derived electron pocket
as a direct result of the enhanced apical distances.
In contrast, tensile strain enhances the occupation of the Nd~$5d_{z^2}$-derived electron pocket
and the Ni~$3d_{z^2}$-derived hole pocket,
paralleling the concomitant depletion of the Ni~$3d_{z^2}$-derived electron pocket.
This is reflected in slight variations of the Ni magnetic moments, which amount to
$\pm 0.93~\mu_\text{B}$ ($a_\text{LAO}$),
$\pm 0.97~\mu_\text{B}$ ($a_\text{STO}$), and
$\pm 0.98~\mu_\text{B}$ ($a_\text{DSO}$),
and demonstrates that the degree of self-doping as well as the electron and hole carrier density can be fine tuned by strain. 
We note that an alternative strategy to epitaxial strain exerted by a substrate may be chemical pressure,
i.e., exploiting the monotonic variation of the $a$ and $c$ lattice parameters across the lanthanide series~\cite{sahinovic_active:21, sahinovic_quantifying:22, NdNiO2-lanthanide-trends:21}.

\section{Summary}

We investigated the magnetic interactions in infinite-layer nickelates versus cuprates by performing first-principles simulations
including a Coulomb repulsion term,
systematically and consecutively varying a number of control parameters
such as the on-site Coulomb and exchange interaction, spin-orbit coupling,
the explicit hole doping, and the treatment of the Nd $4f$ electrons.
The $U$-$J$ phase diagrams for undoped nickelates and cuprates
indicate $G$-type antiferromagnetic (AFM) ordering,
yet with a different $U$ dependence of the magnetic coupling.
By either Sr hole doping or explicit treatment of the Nd $4f$ electrons,
we identified a transition to a Ni $C$-type AFM ground state.
This observation is attributed to a distinct response of
the Ni versus Cu $e_g$ orbitals to the hole doping.
The coupling between Nd $4f$ and Ni $3d$ states stabilizes $C$-type AFM order on both sublattices.
Even though spin-orbit interactions induce a band splitting near the Fermi energy,
the bad-metal state is retained even under epitaxial strain.
These results highlight the unique magnetic interactions present in the infinite-layer nickelates,
which establishes them as an intriguing platform to investigate unconventional superconductivity from a novel perspective.

\begin{acknowledgments}
This work was supported by the German Research Foundation (Deutsche Forschungsgemeinschaft, DFG) 
within the SFB/TRR~80 (Projektnummer 107745057), Project No.~G3, as well as within IRTG 2803 (Projektnummer 461605777),
and the National Science Foundation, Grant No.~NSF-DMR-2118718.
Computing time was granted by the Center for Computational Sciences and Simulation of the University of Duisburg-Essen
(DFG Grants No.~INST 20876/209-1 FUGG and No.~INST 20876/243-1 FUGG).
\end{acknowledgments}

\end{document}